\RequirePackage{fix-cm}

\documentclass[twocolumn,epjc3]{svjour3}

\smartqed  

\RequirePackage{graphicx}
\RequirePackage{amsmath}
\usepackage[stable]{footmisc}	
\usepackage{multirow}	

\RequirePackage{mathptmx}      

\RequirePackage[colorlinks,citecolor=blue,urlcolor=blue,linkcolor=blue]{hyperref}
\usepackage{breakurl} 

\journalname{Eur. Phys. J. C}

\begin{document}

\title{Numerical solutions to Giovannini's parton branching equation up to TeV energies at the LHC
}

\author{Z. Ong\thanksref{e1, addr1} \and
        P. Agarwal\thanksref{addr1} \and
        H.W. Ang\thanksref{addr1} \and
        A.H. Chan\thanksref{addr1} \and
        C.H. Oh\thanksref{addr1}
}

\thankstext{e1}{e-mail: ongzongjin@u.nus.edu}

\institute{Department of Physics, National University of Singapore.	2 Science Drive 3, Singapore 117551.\label{addr1}
}

\date{Received: date / Accepted: date}

\maketitle

\begin{abstract}
Giovannini's parton branching equation is integrated numerically using the 4\textsuperscript{th}-order Runge-Kutta method. Using a simple hadronisation model, a charged-hadron multiplicity distribution is obtained. This model is then fitted to various experimental data up to the TeV scale to study how the Giovannini parameters vary with collision energy and type. The model is able to describe hadronic collisions up to the TeV scale and reveals the emergence of gluonic activity as the centre-of-mass energy increases. A prediction is made for $\sqrt{s}$ = 14 TeV.
\keywords{Multiplicity distribution \and parton branching equation}
\end{abstract}


\section{Introduction}
\label{sec:1}

The charged particle multiplicity distribution is one of the important primary measurements made at every high energy physics collision experiment. Particle physics phenomenology attempts to create models to describe the observed multiplicity distributions, and distributions such as the Negative Binomial Distribution (NBD) and Generalised Multiplicity Distribution (GMD) \cite{Chan1990} have been very successful in describing observed data, at least up to the emergence of KNO scaling violation reported by the UA5 collaboration in the 1980s \cite{Alner1986}. The appearance of a “shoulder-like structure” hinted at a new process that was taking place, and the common approach by phenomenologists since then was to fit the observations to a multi-component distribution. For example, a two-component distribution can be composed as:

\begin{equation}
\label{eq:1}
\begin{split}
P(n;\overline{n},k,\overline{n}',k') = & \alpha P_{\text{NBD}}(n;\overline{n},k) \\
& + (1-\alpha) P_{\text{NBD}}'(n;\overline{n}', k')
\end{split}
\end{equation}

\noindent where the resultant distribution $P(n)$ is a weighted sum of two individual distributions, each with its own set of parameters (see for example ref. \cite{Ghosh2012c}, where the two components describe “soft” and “semihard” events). Three-component distributions have also been explored as a possible explanation to other observed phenomena in multiplicity distributions such as oscillations of combinants \cite{Zborovsky2018}.

Distributions such as the NBD and GMD are partial solutions to Giovannini's parton branching equation \cite{Giovannini1979} (henceforth ``Giovannini's equation") which describes multiparticle production within the framework of perturbative QCD (pQCD). For $m$ quarks and $n$ gluons at QCD evolution parameter $t$, we have:

\begin{equation}
\label{eq:2}
\begin{split}
\frac{\partial P_{m,n}}{\partial t} =
& -(An + \widetilde{A}m + Bn + Cn) P_{m,n} \\
& + A(n-1) P_{m,n-1} \\
& + \widetilde{A} m P_{m,n-1} \\
& + B(n+1) P_{m-2,n+1} \\
& + C(n-2) P_{m,n-2}
\end{split}
\end{equation}

\noindent where $A$, $\widetilde{A}$, $B$ and $C$ (henceforth the “Giovannini parameters”) are the probabilities of the processes

\begin{equation}
\label{eq:3}
\begin{split}
& A: g \rightarrow gg \; &&\text{(gluon bremsstrahlung)} \\
& \widetilde{A}: q \rightarrow qg &&\text{(quark bremsstrahlung)} \\
& B: g \rightarrow q\overline{q} &&\text{(quark pair creation)} \\
& C: g \rightarrow ggg &&\text{(four-gluon vertex)}
\end{split}
\end{equation}

\noindent taking place. The GMD is a partial solution to the parton branching equation for the special case $B = C = 0$, and the NBD is a special case of the GMD where $k' = 0$. It must be noted then, that these solutions do not capture all the processes considered by Giovannini.

At the point of writing, a full analytic distribution incorporating all four of Giovannini's processes does not exist. However, several attempts at a partial solution have been studied and made, for example by Biyajima, Suzuki and Wilk \cite{Biyajima1983,Biyajima1984b,Biyajima1984,Biyajima1984a,Biyajima1988b,Suzuki1991}, Carruthers and Shih \cite{Shih1986,Carruthers1987}, Durand and Sarcevic \cite{Durand1986,Durand1987}, Sakai \cite{Sakai1989}, Gupta and Sarma \cite{Gupta1991}, Hwa \cite{Hwa1988} and Chan and Chew \cite{Chan1992b}. Since an exact solution is difficult to obtain, we look to numerical methods. Of interest to this work is Sakai's attempt at a numerical solution \cite{Sakai1989}; in it, the $C$-process was omitted (and hence a full solution to eq. \eqref{eq:2} was not explored) and best-fit parameters to experimental data were not reported. However, oscillations were present in the hadron multiplicity plots generated (Figures 1(b) and 1(c) in \cite{Sakai1989}), which are of interest and might be relevant to the shoulder-like structure observed in KNO scaling violation.

Hence, with newer data available in the TeV-era of the LHC, we reattempt to construct a “single-component distribution” numerical model to see if it can better describe charged-particle multiplicity distributions, well into the KNO scaling violation regime up to the TeV scale.

This article is organised as follows. Section \ref{sec:2} outlines our numerical model, where the procedure from integrating Giovannini's parton branching equation to applying a hadronisation model to produce the final charged-particle multiplicity distribution is detailed. Section \ref{sec:3} explores some properties of the model parameters and how they affect the final multiplicity distribution. Section \ref{sec:4} outlines the fitting procedure used to fit against data and the results, followed by a discussion about it in section \ref{sec:5}. In section \ref{sec:6} we make a prediction for higher energies, and we end in section \ref{sec:7} with some concluding remarks.


\section{Our model}
\label{sec:2}

Our model for generating a charged-particle multiplicity distribution is adapted from Sakai \cite{Sakai1989} with a few key extensions. There are four stages to this model: (1) parton branching, (2) hadronisation, (3) selecting charged final states from all hadrons produced and (4) an optional stage to account for particles produced but not detected due to reduced detector phase space.


\subsection*{Stage 1: parton branching}
\label{sec:2.1}

A two-dimensional probability distribution $P[m][n]$, where $m$ = no. of quarks and $n$ = no. of gluons is first set up. For an initial condition of $M=2$ quarks and $N=0$ gluons as an example, we initialise the array with $P[2][0]=1$ and all other $P[m][n]=0$. Each element of $P[m][n]$ is then integrated with eq. \eqref{eq:2} using the RK4 routine with step size $h = 0.01$, for $t = 0.0$ to an upper bound $T$. The Giovannini parameters $A$, $\widetilde{A}$, $B$ and $C$ take on a real number between 0 and 1, with $A+B+C\leq1$ due to conservation of probability. After that, $P[m][n]$ is condensed into a one-dimensional parton distribution $P_{\text{parton}}(n)$, where quarks and gluons are categorised together. $P_{\text{parton}}(n)$ is truncated at a maximum value $n=n_{\text{max}}$ (from the data fitted against) and renormalised to unity.

In our analysis, $A$, $\widetilde{A}$, $B$, $C$, $M$ and $N$ will be free parameters that we will optimise for each dataset. $T$ will be kept constant at 3.0 because it was found that it varies approximately inversely to $A$, $\widetilde{A}$, $B$ and $C$ (i.e. doubling $T$ approximately halves $A$, $\widetilde{A}$, $B$ and $C$). It is also important to note that the Giovannini parameters are held constant throughout, which reflects a self-similarity mechanism in the parton branching process. Additionally, these constants would be determined by the nature and centre-of-mass energies of the collisions.


\subsection*{Stage 2: hadronisation}
\label{sec:2.2}

After parton branching terminates, each parton hadronises into $l$ hadrons according to a modified Poisson distribution $H(l)$, parametrised with average $l_{\text{avg}}$ and up to a maximum of $l_{\text{max}}$ hadrons. $H(l)$ is a subset of the usual Poisson distribution, with $l$ taking integer values between 1 and $l_{\text{max}}$. 

\begin{equation}
\label{eq:4}
H(l) = \frac{f(l_{\text{avg}},l)}{\sum_{l=1}^{l_\text{max}}f(l_{\text{avg}},l)}
\end{equation}

\noindent where $f(l_{\text{avg}},l) = \frac{\exp(-l_{\text{avg}})\cdot(l_{\text{avg}})^l}{l!}$ is the usual probability mass function of the Poisson distribution.

This hadronisation step applies identically to both quarks and gluons. After hadronisation, we end up with an all-hadron multiplicity distribution $P_{\text{all-hadron}}(n)$, which includes both charged and neutral hadrons. In our analysis, $l_{\text{avg}}$ and $l_{\text{max}}$ will be free parameters that we will optimise for each dataset. In the special case where the best-fit value of $l_{\text{max}} = 1$, we have 1-to-1 hadronisation (one parton becomes one hadron) regardless of the value of $l_{\text{avg}}$.


\subsection*{Stage 3: selecting charged final states}
\label{sec:2.3}

To obtain a multiplicity distribution for only the charged final states, we assume the same process as described in ref. \cite{Sakai1989}: the initial state begins with an even number of charged particles, resulting in an overall charge of 0 (corresponding to $e^{+}e^{-}$ and $\overline{p}p$ collisions) or 2 (corresponding to $pp$ collisions). To conserve electric charge, all the resultant particles produced can be group into either $(+,-)$ or $(0,0)$ pairs. If there are an odd number of hadrons produced, the last hadron will be neutral.

We assume that each parton has an equal probability of hadronising into a charged hadron and a neutral hadron. Hence, the probability that $2l$ hadrons contain $2l_{c}$ charged hadrons is given by a combinatorial formula:

\begin{equation}
\label{eq:5}
\begin{split}
P_{\text{charged-hadron}}(2l_{c}) = &\sum_{l\geq l_{c}} \frac{l!}{l_{c}!(l-l_{c})!2^{l}} [P_{\text{all-hadron}}(2l) \\
&+ P_{\text{all-hadron}}(2l+1)]
\end{split}
\end{equation}

Eq. \eqref{eq:5} gives us the charged particle multiplicity distribution for full phase space and is nonzero only for even multiplicities. From here on, $P_{\text{charged-hadron}}(n)$ will be referred to as simply $P(n)$.


\subsection*{Stage 4: reducing pseudorapidity interval}
\label{sec:2.4}

Finally, we note that some published data only represent a subset of all charged particles produced. These can be due to various factors, such as track selection criteria or constraints due to detector design. For example, the CMS collaboration \cite{Khachatryan2011a} at CERN only presents charged-particle multiplicity data for particles produced within pseudorapidity range $|\eta|\leq 2.4$ due to detector geometry.

To account for particles that are not counted due to the limited pseudorapidity intervals of the data, we implement a simple "loss function" adapted from another work by one of the authors (P. Agarwal)\footnote{To appear in proceedings of SEAAN Meeting 2019.}. For each particle produced, there is a probability $p_\text{loss}$ that it does not get detected or included in the data. We propose a Binomial distribution to model this loss function; for $n$ charged particles produced, the probability $\tilde{p}$ of losing $i$ particles is

\begin{equation}
\label{eq:6}
\tilde{p}(n,i) = \frac{n!}{i!(n-i)!} (p_{\text{loss}})^i (1 - p_{\text{loss}})^{n-i}
\end{equation}

Hence, $P_{\text{reduced}}(n)$ for reduced phase space will be given by

\begin{equation}
\label{eq:7}
P_{\text{reduced}}(n) = \sum_{i} P(n+i) \cdot \tilde{p}(n+i,i)
\end{equation}

The loss function will be applied in fits to data with reduced phase space in section \ref{sec:4}.


\section{Some properties of the model\footnote{This section is an extension of a prior preliminary work by the author (Z. Ong) to appear in proceedings of SEAAN Meeting 2019.}}
\label{sec:3}

We now explore the behaviour of the tunable parameters in the model. In each plot that follows, three contrasting values of a selected parameter are shown to illustrate the effect they have on the eventual multiplicity distribution. Apart from one selected parameter in each plot, all parameters are set to an arbitrarily chosen set of values: $A = 0.2$, $\widetilde{A} = 0.2$, $B = 0$, $C = 0$, $M = 2$, $N = 0$, $l_{\text{avg}} = 1.0$, $l_{\text{max}} = 5$ and $T = 3.0$. Stage 4 of the model (reducing phase space) will only be applied in Figure \ref{fig:vary_ploss}. In all generated multiplicity distributions, the number of charged particles will be capped at 50 and renormalised to 1.

\begin{figure}
	\includegraphics[width=1.1\columnwidth]{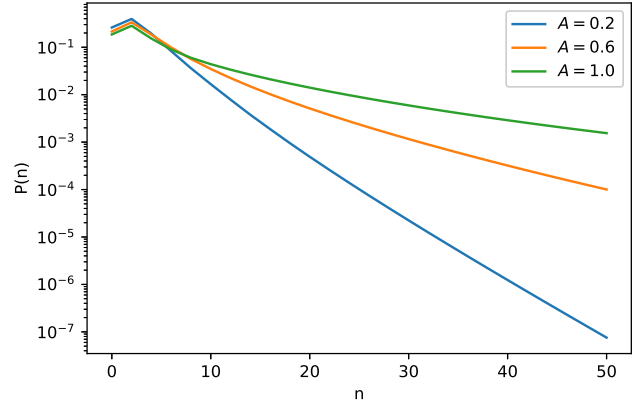}
	\caption{Varying $A$}
	\label{fig:vary_A}       
\end{figure}

\begin{figure}
	\includegraphics[width=1.1\columnwidth]{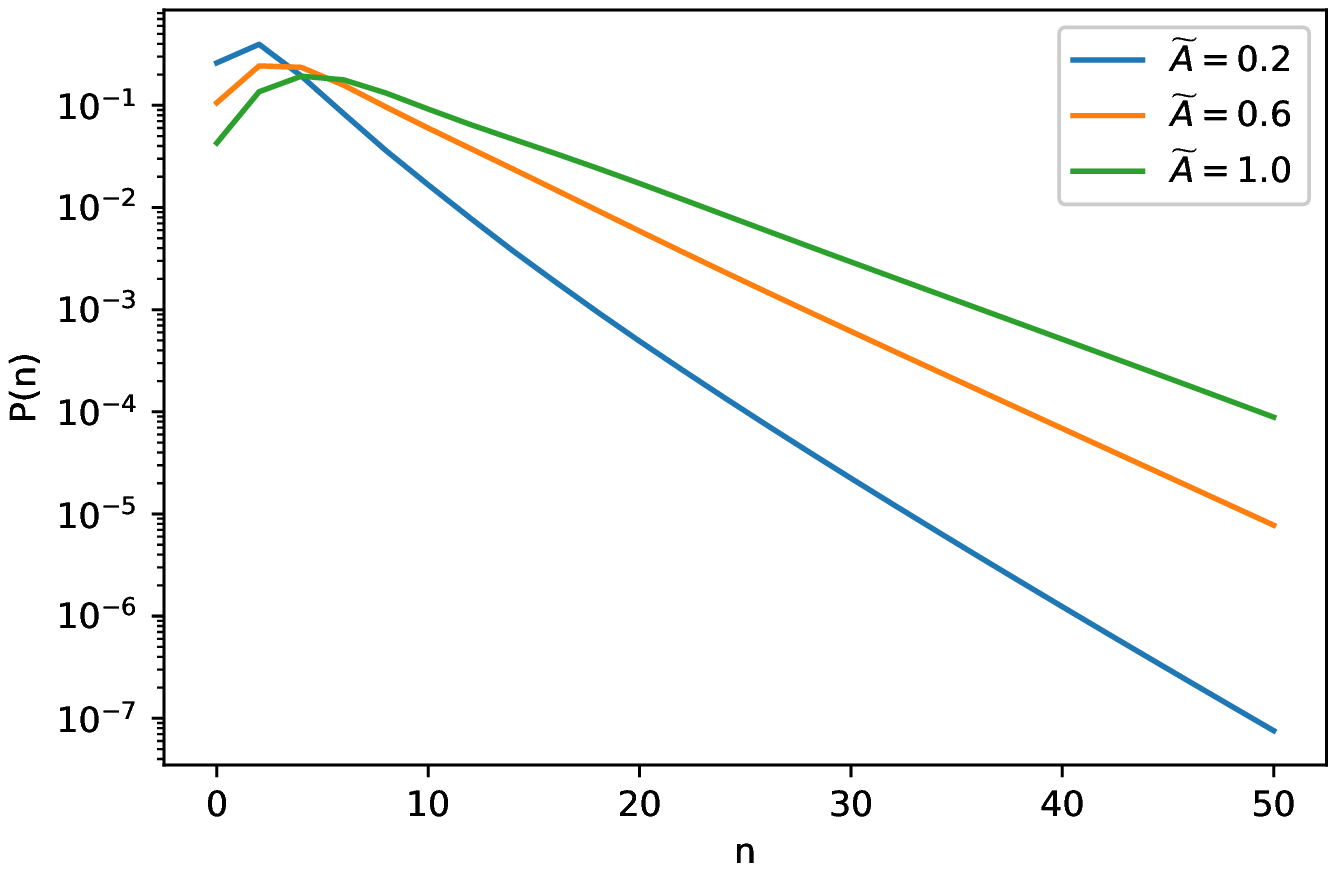}
	\caption{Varying $\widetilde{A}$}
	\label{fig:vary_TA}
\end{figure}

\begin{figure}
	\includegraphics[width=1.1\columnwidth]{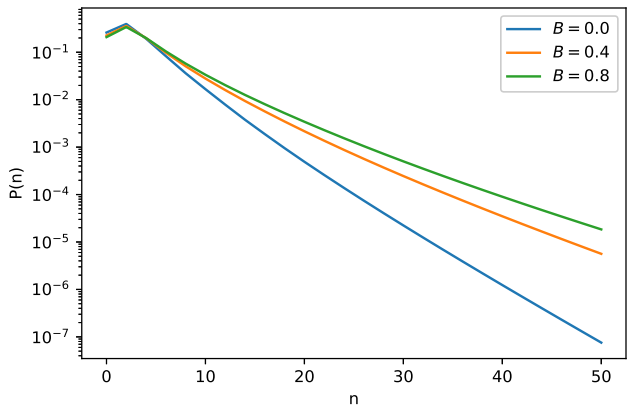}
	\caption{Varying $B$}
	\label{fig:vary_B}
\end{figure}

\begin{figure}
	\includegraphics[width=1.1\columnwidth]{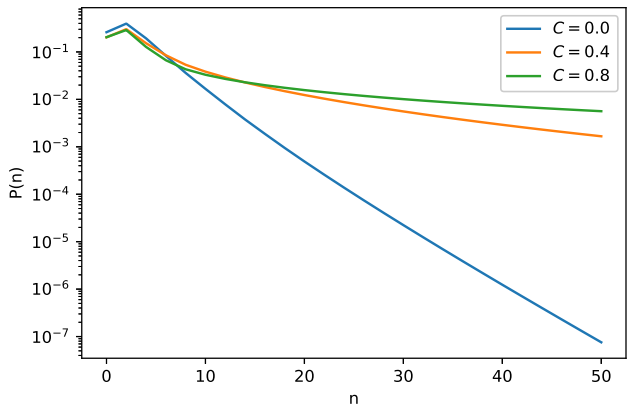}
	\caption{Varying $C$}
	\label{fig:vary_C}
\end{figure}

\begin{figure}
	\includegraphics[width=1.1\columnwidth]{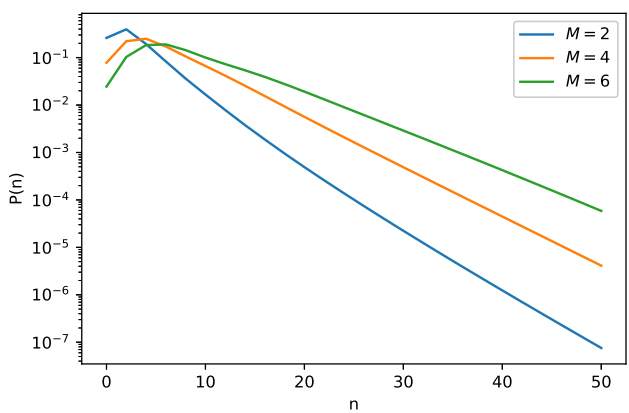}
	\caption{Varying $M$}
	\label{fig:vary_M}
\end{figure}

\begin{figure}
	\includegraphics[width=1.1\columnwidth]{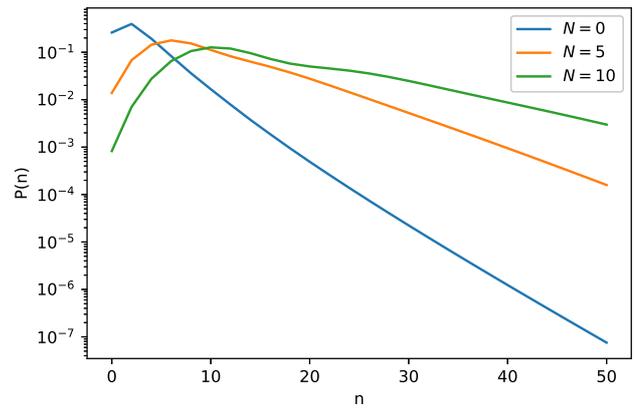}
	\caption{Varying $N$}
	\label{fig:vary_N}
\end{figure}

\begin{figure}
	\includegraphics[width=1.1\columnwidth]{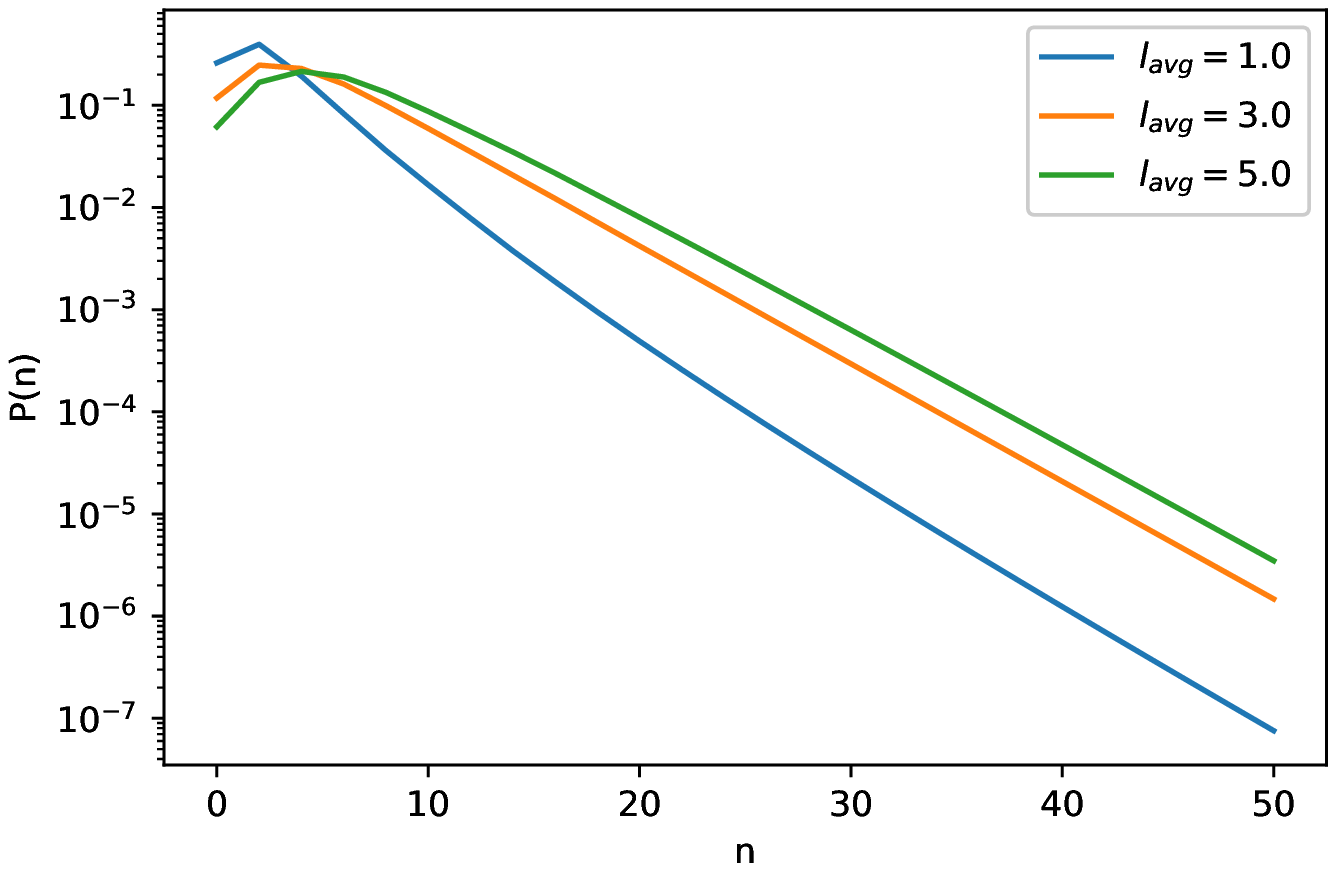}
	\caption{Varying $l_{\text{avg}}$}
	\label{fig:vary_lavg}
\end{figure}

\begin{figure}
	\includegraphics[width=1.1\columnwidth]{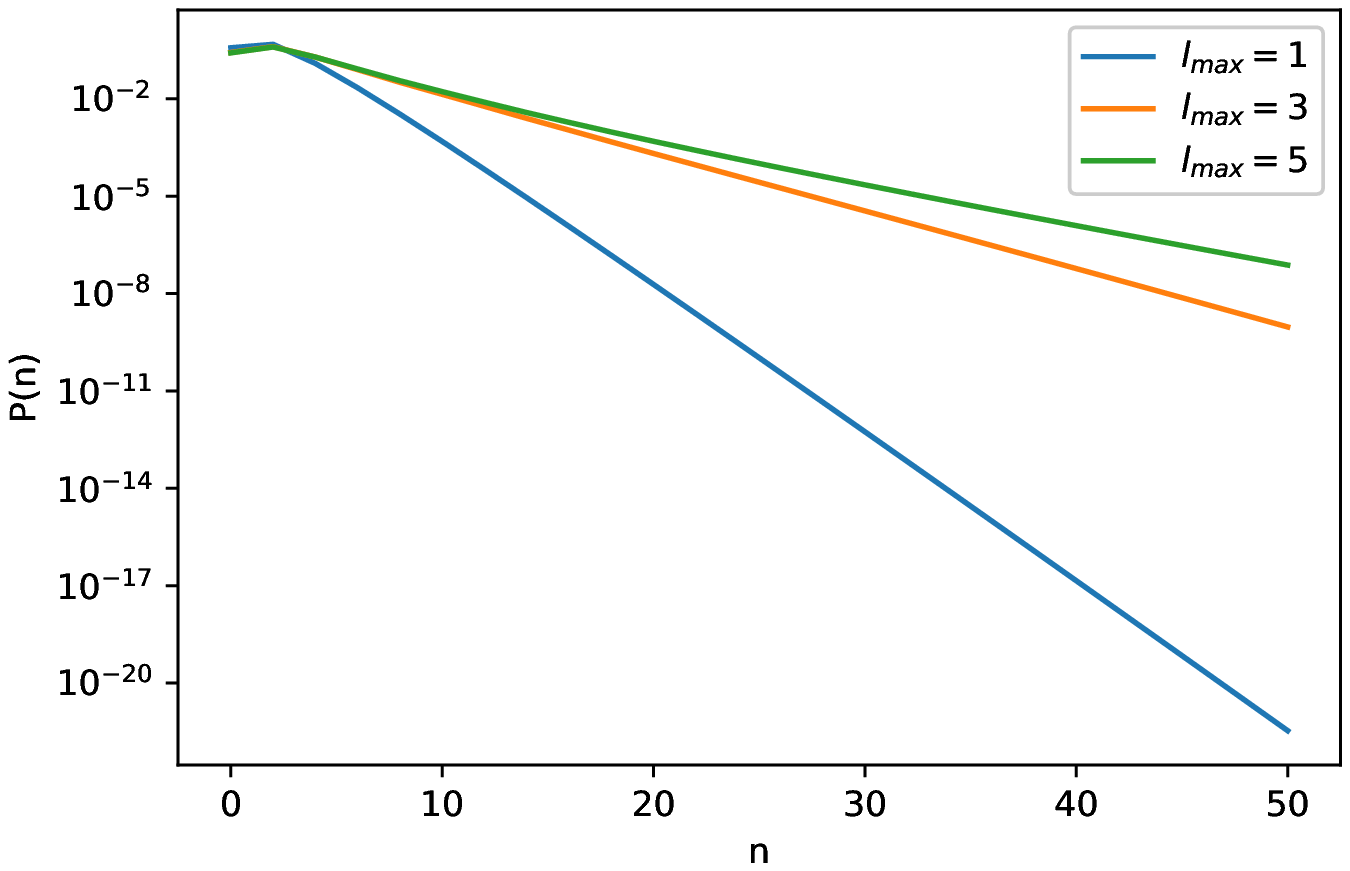}
	\caption{Varying $l_{\text{max}}$}
	\label{fig:vary_lmax}
\end{figure}

\begin{figure}
	\includegraphics[width=1.1\columnwidth]{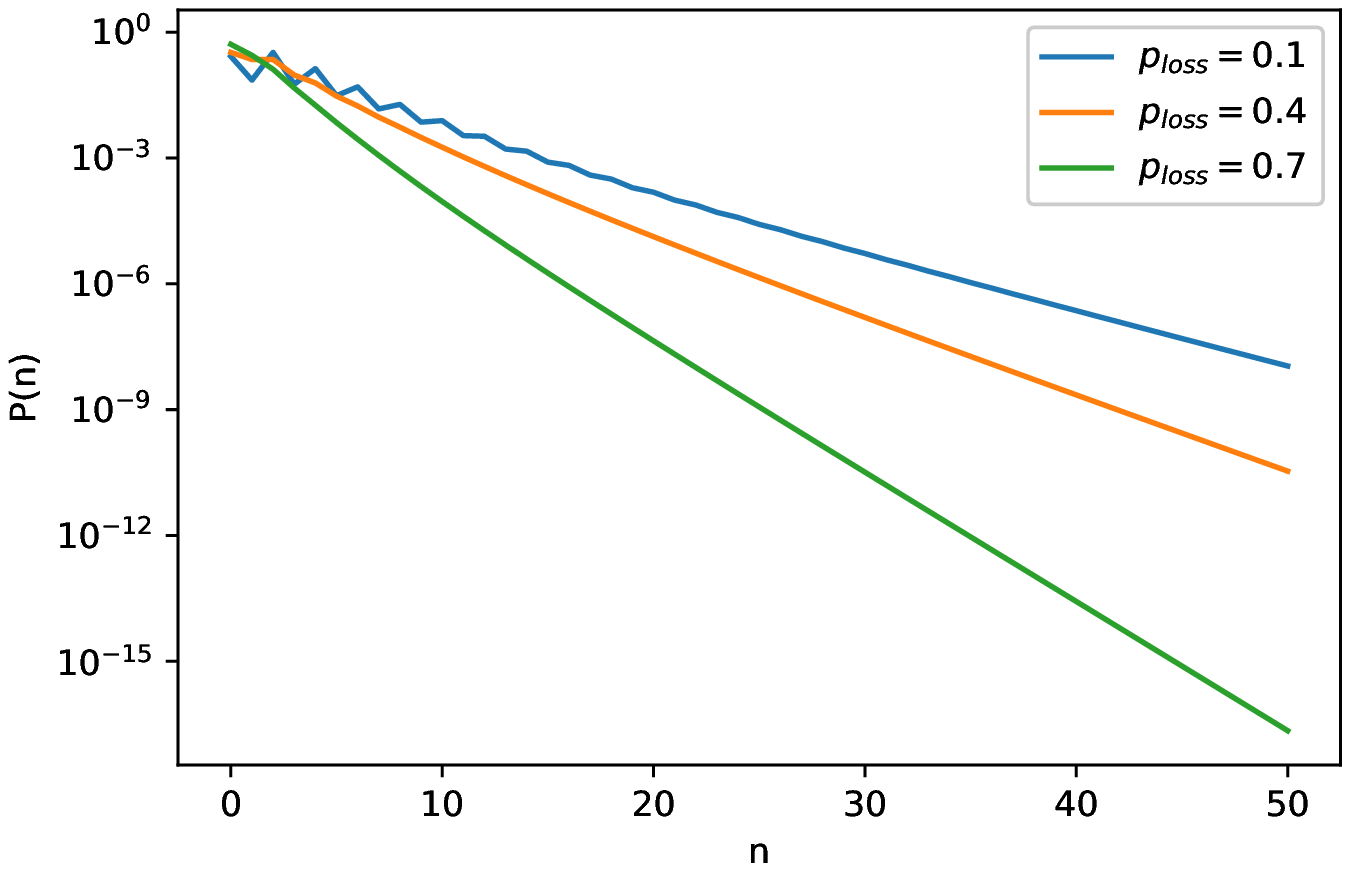}
	\caption{Varying $p_{\text{loss}}$}
	\label{fig:vary_ploss}
\end{figure}

Higher values of each parameter generally have an effect of broadening the overall distribution. However, some parameters exhibit different behaviour from the rest, and we note several interesting observations below.

Figures \ref{fig:vary_A} and \ref{fig:vary_C} show that high values of $A$ or $C$ produce a noticeable change in gradient at approximately $n=5$. This is very similar to the shoulder-like structure observed in KNO scaling violation, mostly evident in hadronic collisions above $\sqrt{s} = 200$ GeV (see Figures \ref{fig:Fits_UA5_ppbar} to \ref{fig:Fits_CMS_7000}). These correspond to gluon branching processes, which suggests they may play an important role in the physics behind KNO scaling violation.

Figures \ref{fig:vary_M} and \ref{fig:vary_N} show that the initial number of quarks ($M$) and gluons ($N$) in the collision affects the position of the distribution's peak. However, the extremely high values (such as $M = 10$ in Figure \ref{fig:vary_N}) might introduce oscillations in the final distribution.

Figure \ref{fig:vary_ploss} shows a potential weakness of the Binomial loss function. At low values (see $p_{\text{loss}} = 0.1$ in Figure \ref{fig:vary_ploss}), oscillations occur at multiplicities below $n = 20$, reflective of numerical calculation instability. Hence, we would not expect best-fit values of $p_{\text{loss}}$ that are too low.

Together, the nine parameters bring about some effect on the eventual shape of the multiplicity distribution. However, not all effects are independent (e.g. parameters $A$ and $C$); hence, it is certainly plausible that two different sets of parameter values can result in equally good fits to the experimental data. In this work, we limit our scope to demonstrating the ability of the numerical model to describe the data – showing the existence and derivation of comparable solution sets akin to ``family of curves" shall be considered in a future work.


\section{Fits to experimental data}
\label{sec:4}

In this section, we attempt to fit the multiplicity distribution produced by our numerical model against a selection of experimental data: $e^{+}e^{-}$ annihilations at $\sqrt{s} =$ 14, 22, 34.8 and 43.6 GeV from TASSO \cite{TASSOCollaboration1989}, $\sqrt{s} =$ 57 GeV from AMY \cite{TheAMYCollaboration1990}, $\sqrt{s} =$ 91 GeV from DELPHI \cite{Abreu1991} and $\sqrt{s} =$ 133, 161, 172, 183 and 189 GeV from OPAL \cite{OPALcollaboration1996,OPALcollaboration1997,Ackerstaff1997a}; $p\overline{p}$ collisions at $\sqrt{s} =$ 200, 546 and 900 GeV from UA5 \cite{Alner1987,UA5Collaboration1989}, and $pp$ collisions at $\sqrt{s} =$ 0.9, 2.36 and 7 TeV from CMS \cite{Khachatryan2011a}. For CMS data, the data point at $n = 0$ is omitted.

\begin{figure*}
	\includegraphics[width=0.75\textwidth]{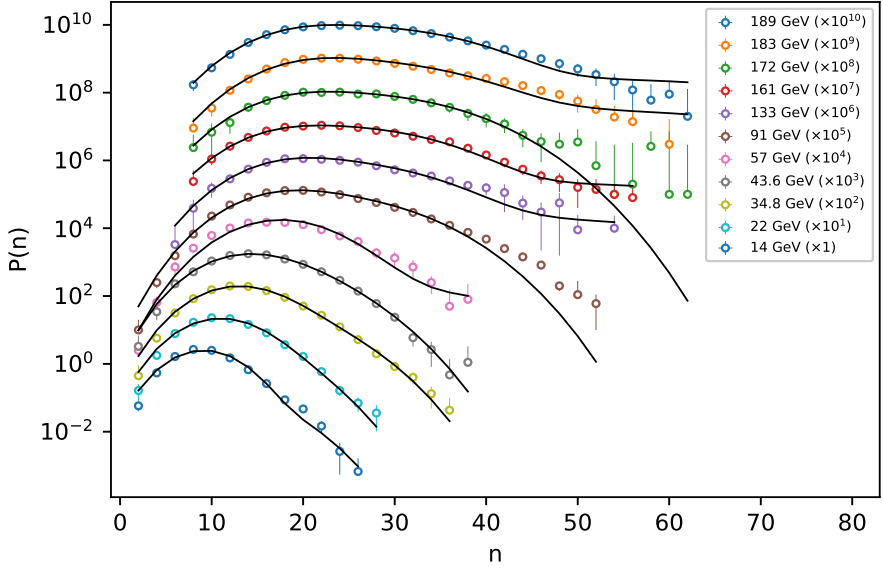}
	\caption{Fits to $e^{+}e^{-}$ collisions (various collaborations)}
	\label{fig:Fits_e+e-}
\end{figure*}

\begin{figure*}
	\includegraphics[width=0.75\textwidth]{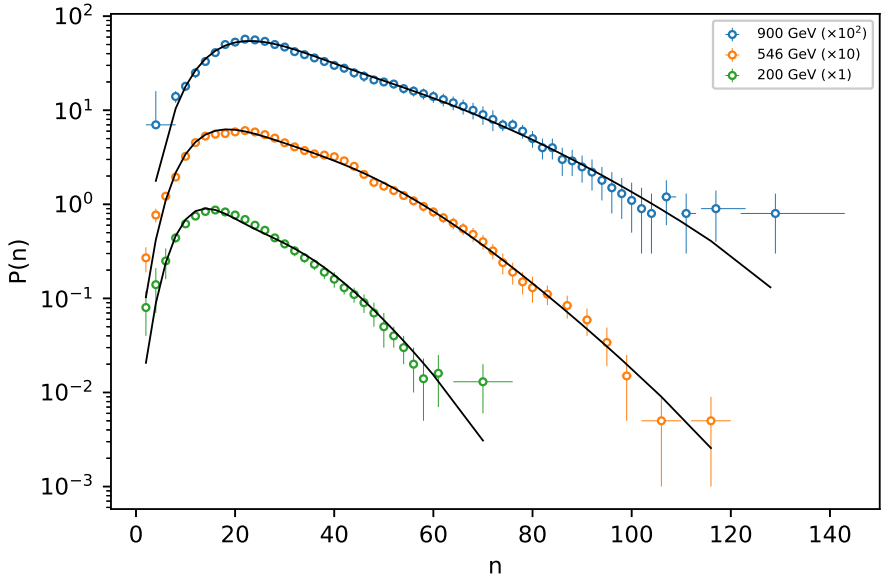}
	\caption{Fits to $\overline{p}p$ collisions (UA5 collaborations)}
	\label{fig:Fits_UA5_ppbar}
\end{figure*}

\begin{figure*}
	\includegraphics[width=0.75\textwidth]{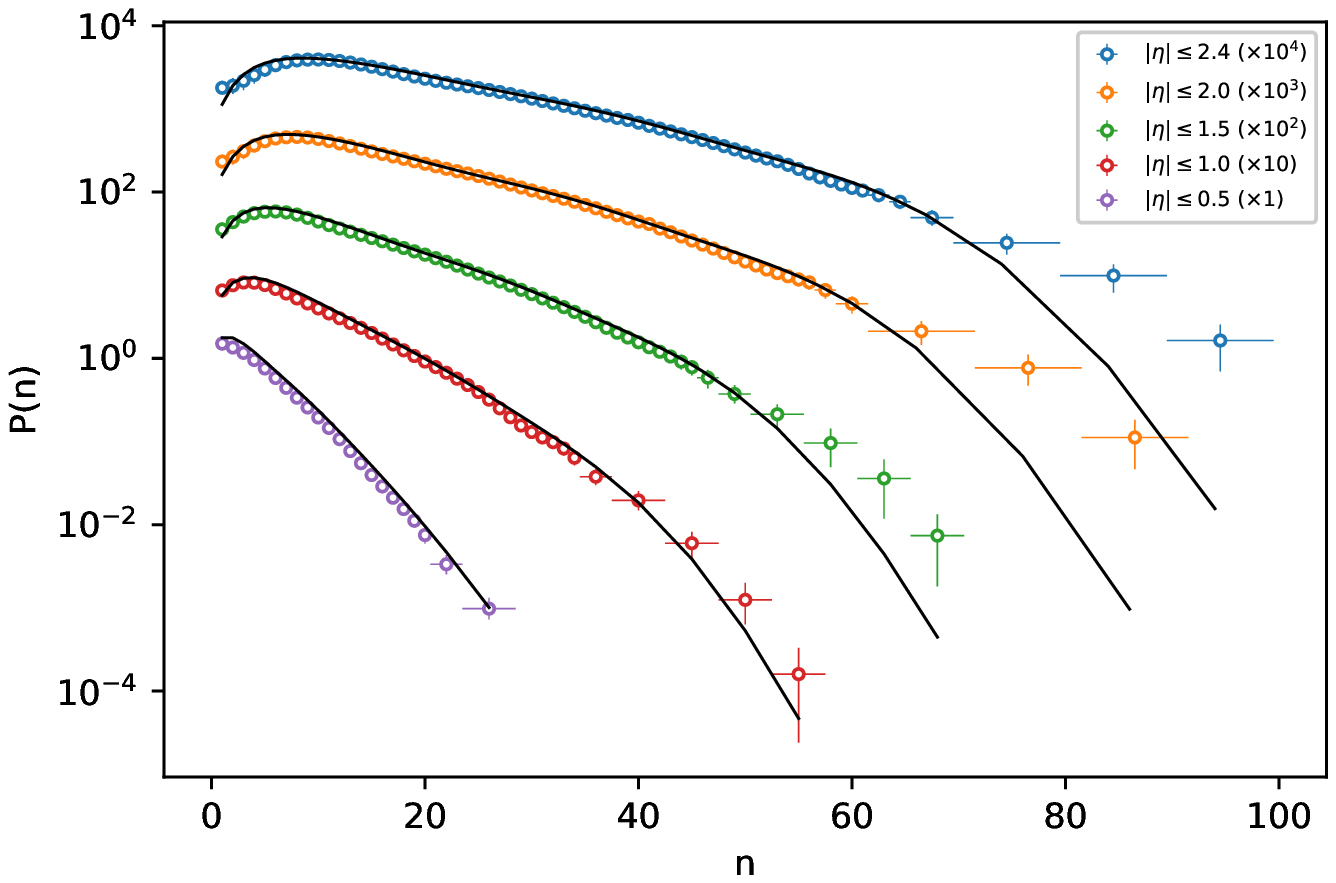}
	\caption{Fits to $pp$ collisions at $\sqrt{s} =$ 0.9 TeV (CMS collaboration)}
	\label{fig:Fits_CMS_900}
\end{figure*}

\begin{figure*}
	\includegraphics[width=0.75\textwidth]{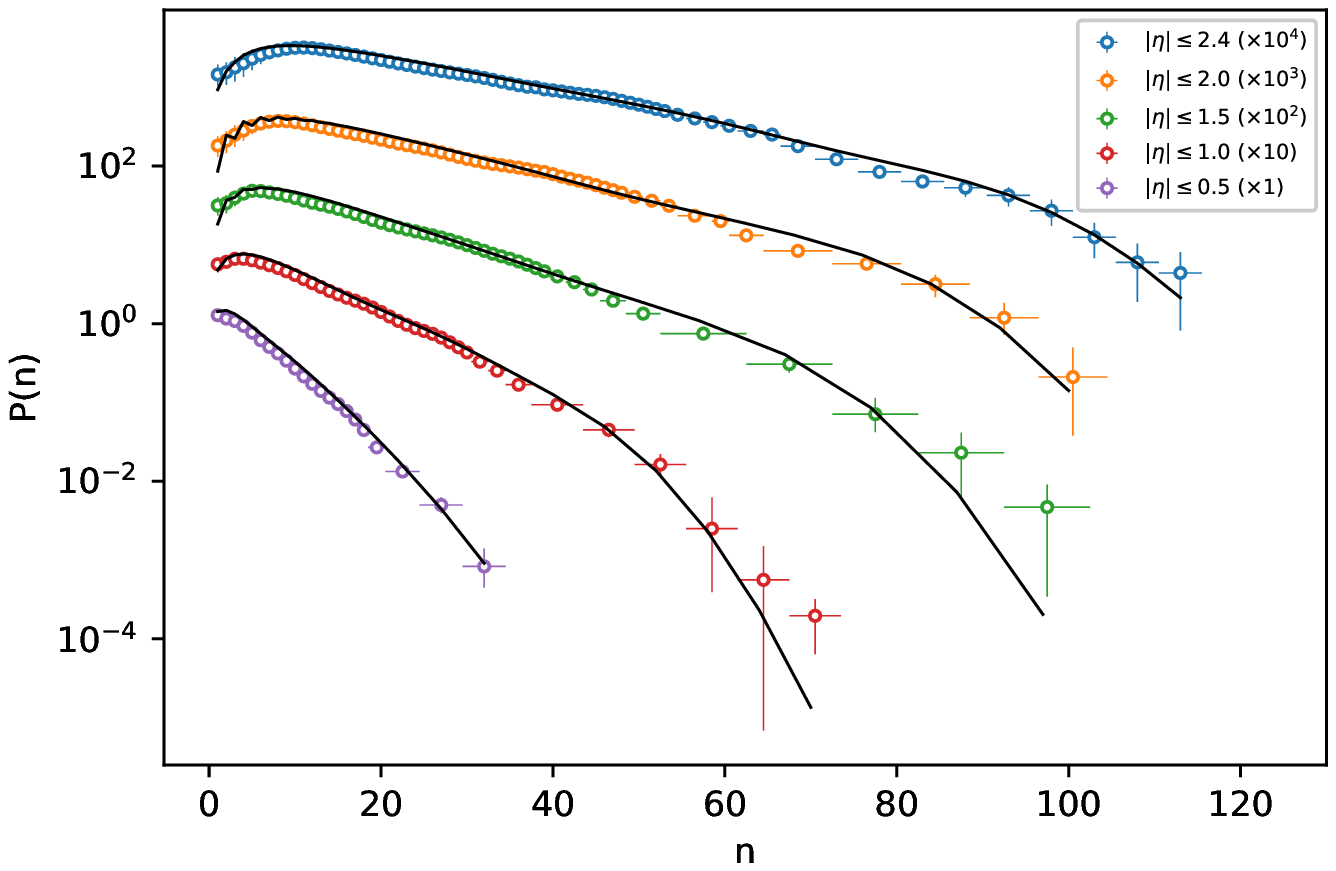}
	\caption{Fits to $pp$ collisions at $\sqrt{s} =$ 2.36 TeV (CMS collaboration)}
	\label{fig:Fits_CMS_2360}
\end{figure*}

\begin{figure*}
	\includegraphics[width=0.75\textwidth]{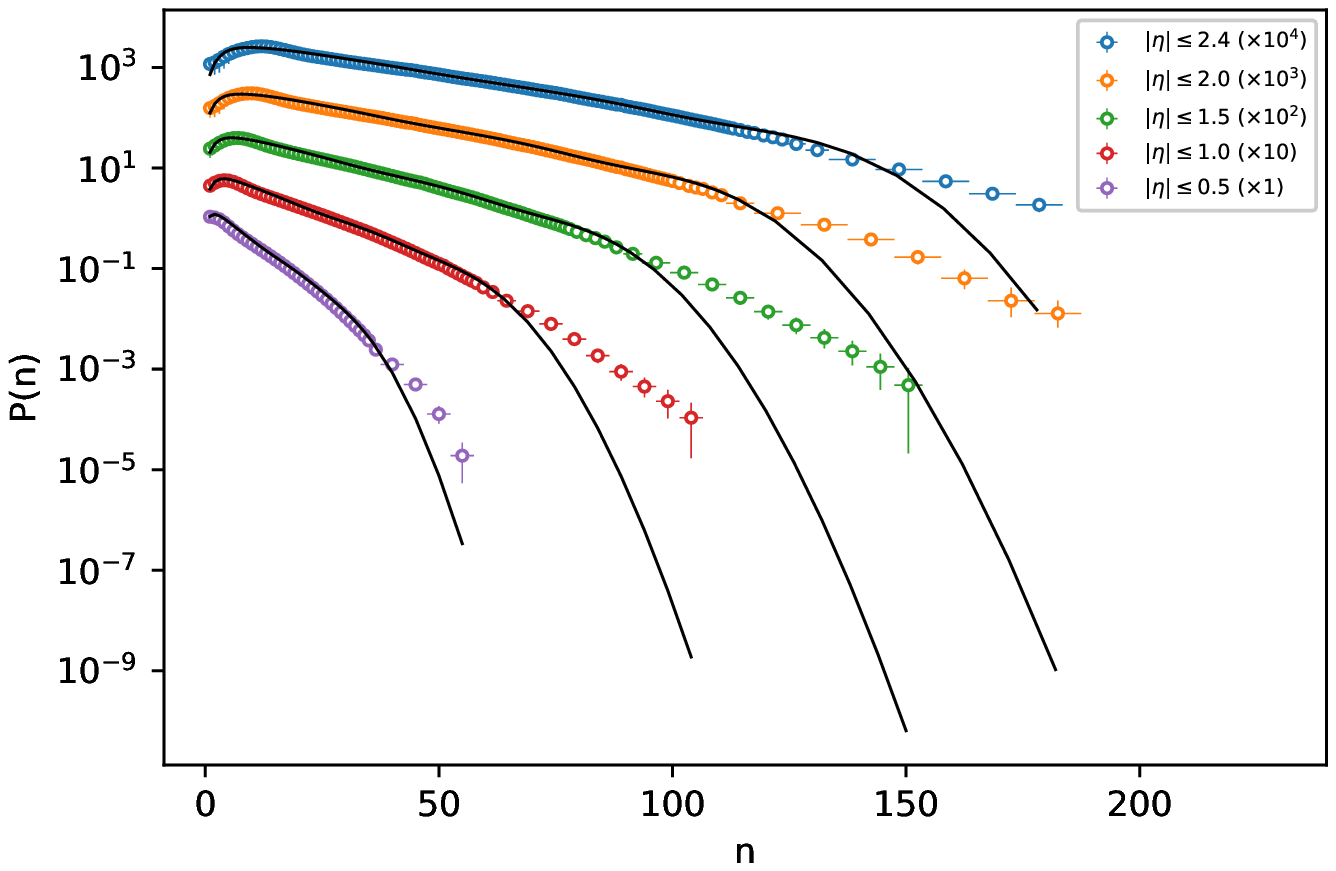}
	\caption{Fits to $pp$ collisions at $\sqrt{s} =$ 7 TeV (CMS collaboration)}
	\label{fig:Fits_CMS_7000}
\end{figure*}

The best-fit values for the model parameters are searched using a Markov chain Monte Carlo (MCMC) algorithm and are presented in Tables \ref{tab:1a}-\ref{tab:1c} for the three types of collisions. Figures \ref{fig:Fits_e+e-}-\ref{fig:Fits_CMS_7000} show semi-log plots of the probability distributions that are measured experimentally (circles with error bars) and generated by our model (solid lines). Finally, to further compare the distributions from data and our model, we give the normalised moments $C_{q} = \langle n^{q} \rangle / \langle n \rangle ^{q}$ for $q =$ 2-5 in Tables \ref{tab:2a}-\ref{tab:2e}.

\begin{table*}[]
	\caption{Best-fit parameters ($e^{+}e^{-}$ collisions, various collaborations)}
	\label{tab:1a}
	\centering
	\begin{tabular}{ccccccccccc}
		\hline
		$\sqrt{s}$ (GeV) & Collab. & $A$ & $\widetilde{A}$ & $B$ & $C$ & $M$ & $N$ & $l_{\text{avg}}$ & $l_{\text{max}}$ & $\chi^{2}$/d.o.f.\\
		\hline
		14   & TASSO  & 0.0647 & 0.0047 & 0.1164 & 0.3512 & 18 & 0 & 0.8793 & 1 & 16.5642 \\
		22   & TASSO  & 0.0135 & 0.0022 & 0.1682 & 0.1659 & 20 & 1 & 0.9492 & 1 & 8.6087  \\
		34.8 & TASSO  & 0.1370 & 0.0135 & 0.3716 & 0.2929 & 25 & 0 & 0.3910 & 1 & 6.9343  \\
		43.6 & TASSO  & 0.0175 & 0.0006 & 0.4582 & 0.1048 & 11 & 8 & 0.6484 & 1 & 1.0916  \\
		57   & AMY    & 0.2833 & 0.0081 & 0.0789 & 0.3993 & 34 & 0 & 0.0173 & 3 & 2.7237  \\
		91   & DELPHI & 0.3131 & 0.0022 & 0.3513 & 0.3198 & 30 & 2 & 0.8385 & 1 & 5.3986  \\
		133  & OPAL   & 0.1143 & 0.0028 & 0.2827 & 0.5884 & 32 & 2 & 0.0666 & 5 & 0.5988  \\
		161  & OPAL   & 0.4308 & 0.0008 & 0.1344 & 0.4303 & 31 & 2 & 0.0746 & 4 & 0.4484  \\
		172  & OPAL   & 0.1383 & 0.0079 & 0.4149 & 0.4334 & 30 & 3 & 0.4580 & 1 & 0.1392  \\
		183  & OPAL   & 0.0340 & 0.0001 & 0.1859 & 0.6935 & 37 & 2 & 0.1220 & 2 & 1.4377  \\
		189  & OPAL   & 0.0125 & 0.0097 & 0.1622 & 0.7508 & 34 & 2 & 0.1013 & 2 & 0.8462  \\
		\hline
	\end{tabular}
\end{table*}

\begin{table*}[]
	\caption{Best-fit parameters ($p\overline{p}$ collisions, UA5 collaboration)}
	\label{tab:1b}
	\centering
	\begin{tabular}{cccccccccc}
		\hline
		$\sqrt{s}$ (GeV) & $A$ & $\widetilde{A}$ & $B$ & $C$ & $M$ & $N$ & $l_{\text{avg}}$ & $l_{\text{max}}$ & $\chi^{2}$/d.o.f.\\
		\hline
		200 & 0.0671 & 0.1918 & 0.0842 & 0.1269 & 6 & 6 & 1.4672 & 2 & 0.9352 \\
		546 & 0.2623 & 0.0607 & 0.0279 & 0.1115 & 0 & 9 & 1.9908 & 2 & 1.0409 \\
		900 & 0.3073 & 0.4182 & 0.6328 & 0.0591 & 9 & 0 & 1.3253 & 2 & 0.3276 \\
		\hline
	\end{tabular}
\end{table*}

\begin{table*}[]
	\caption{Best-fit parameters ($pp$ collisions, CMS collaboration)}
	\label{tab:1c}
	\centering
	\begin{tabular}{cccccccccccc}
		\hline
		$\sqrt{s}$ (GeV) & $|\eta| \leq$ & $A$ & $\widetilde{A}$ & $B$ & $C$ & $M$ & $N$ & $l_{\text{avg}}$ & $l_{\text{max}}$ & $p_{\text{loss}}$ & $\chi^{2}$/d.o.f.\\
		\hline
		\multirow{5}{*}{900}  & 0.5 & 0.1829 & 0.2592 & 0.0134 & 0.6120 & 1 & 2 & 1.5059 & 7 & 0.7306 & 0.2754 \\
		& 1.0 & 0.3430 & 0.3457 & 0.0016 & 0.1073 & 0 & 4 & 1.7684 & 3 & 0.5738 & 0.6614 \\
		& 1.5 & 0.1527 & 0.0944 & 0.0368 & 0.2514 & 1 & 4 & 2.0122 & 3 & 0.5594 & 0.3717 \\
		& 2.0 & 0.0561 & 0.3101 & 0.0089 & 0.3211 & 0 & 5 & 1.9857 & 3 & 0.5586 & 0.4550 \\
		& 2.4 & 0.0348 & 0.1196 & 0.0148 & 0.3448 & 0 & 5 & 2.5459 & 3 & 0.5455 & 0.5009 \\
		\hline
		\multirow{5}{*}{2360} & 0.5 & 0.1893 & 0.4830 & 0.0497 & 0.7593 & 2 & 1 & 1.7598 & 7 & 0.7320 & 0.6909 \\
		& 1.0 & 0.0606 & 0.0419 & 0.0242 & 0.2911 & 0 & 4 & 1.8010 & 3 & 0.5681 & 0.6545 \\
		& 1.5 & 0.0023 & 0.0008 & 0.0534 & 0.3312 & 0 & 4 & 1.9608 & 2 & 0.2805 & 1.1752 \\
		& 2.0 & 0.0124 & 0.0204 & 0.0585 & 0.3523 & 0 & 4 & 1.9992 & 2 & 0.1759 & 1.1930 \\
		& 2.4 & 0.1171 & 0.3178 & 0.0194 & 0.3314 & 0 & 4 & 2.8365 & 3 & 0.4358 & 0.4126 \\
		\hline
		\multirow{5}{*}{7000} & 0.5 & 0.0061 & 0.0002 & 0.2361 & 0.3473 & 0 & 3 & 2.0296 & 3 & 0.6302 & 1.5980 \\
		& 1.0 & 0.0433 & 0.0044 & 0.4275 & 0.4280 & 0 & 4 & 2.2061 & 3 & 0.6200 & 1.6879 \\
		& 1.5 & 0.0477 & 0.0016 & 0.2680 & 0.4598 & 0 & 4 & 2.4886 & 3 & 0.6132 & 1.5041 \\
		& 2.0 & 0.0786 & 0.0062 & 0.2400 & 0.4769 & 0 & 4 & 2.6558 & 3 & 0.5886 & 1.1997 \\
		& 2.4 & 0.1859 & 0.0135 & 0.2934 & 0.4120 & 0 & 4 & 2.9969 & 3 & 0.4793 & 1.0158 \\
		\hline
	\end{tabular}
\end{table*}

\begin{table*}[]
	\caption{Moments for $e^{+}e^{-}$ fits}
	\label{tab:2a}
	\centering
	\begin{tabular}{c|ccc|ccc|ccc}
		\hline
		& \multicolumn{3}{c|}{$\sqrt{s}=$ 14 GeV}     & \multicolumn{3}{c|}{$\sqrt{s}=$ 22 GeV}     & \multicolumn{3}{c}{$\sqrt{s}=$ 34.8 GeV}      \\
		& Data    & Model   & \% diff. & Data    & Model   & \% diff. & Data    & Model   & \% diff. \\
		\hline
		$C_{2}$   & 1.1092 & 1.1169 & 0.6887 & 1.0983  & 1.1050  & 0.6067 & 1.0929  & 1.0960  & 0.2890 \\
		$C_{3}$   & 1.3483 & 1.3608 & 0.9192 & 1.3095  & 1.3239  & 1.0950 & 1.2925  & 1.2994  & 0.5344 \\
		$C_{4}$   & 1.7833 & 1.7863 & 0.1697 & 1.6813  & 1.7008  & 1.1529 & 1.6429  & 1.6533  & 0.6278 \\
		$C_{5}$   & 2.5537 & 2.5102 & 1.7181 & 2.3104  & 2.3254  & 0.6500 & 2.2328  & 2.2451  & 0.5488 \\
		$\overline{n}$ & 9.3021 & 9.2921 & 0.1074 & 11.3076 & 11.3077 & 0.0010 & 13.5870 & 13.5800 & 0.0515 \\
		\hline
	\end{tabular}

	\begin{tabular}{c|ccc|ccc|ccc}
		\hline
		& \multicolumn{3}{c|}{$\sqrt{s}=$ 43.6 GeV}     & \multicolumn{3}{c|}{$\sqrt{s}=$ 57 GeV}     & \multicolumn{3}{c}{$\sqrt{s}=$ 91 GeV}      \\
		& Data    & Model   & \% diff. & Data    & Model   & \% diff. & Data    & Model   & \% diff. \\
		\hline
		$C_{2}$   & 1.0928  & 1.0941  & 0.1153 & 1.0883  & 1.0670  & 1.9758  & 1.0922  & 1.0837  & 0.7812 \\
		$C_{3}$   & 1.2892  & 1.2917  & 0.1926 & 1.2759  & 1.2065  & 5.5876  & 1.2931  & 1.2602  & 2.5755 \\
		$C_{4}$   & 1.6272  & 1.6304  & 0.1997 & 1.5984  & 1.4402  & 10.4107 & 1.6483  & 1.5598  & 5.5141 \\
		$C_{5}$   & 2.1818  & 2.1849  & 0.1405 & 2.1251  & 1.8108  & 15.9699 & 2.2469  & 2.0402  & 9.6455 \\
		$\overline{n}$ & 15.0775 & 15.0877 & 0.0675 & 17.4905 & 17.9617 & 2.6578  & 21.1974 & 21.2994 & 0.4798 \\
		\hline
	\end{tabular}

	\begin{tabular}{c|ccc|ccc|ccc}
		\hline
		& \multicolumn{3}{c|}{$\sqrt{s}=$ 133 GeV}     & \multicolumn{3}{c|}{$\sqrt{s}=$ 161 GeV}     & \multicolumn{3}{c}{$\sqrt{s}=$ 172 GeV}      \\
		& Data    & Model   & \% diff. & Data    & Model   & \% diff. & Data    & Model   & \% diff. \\
		\hline
		$C_{2}$   & 1.1045  & 1.0957  & 0.7944 & 1.0976  & 1.0972  & 0.0302 & 1.0899  & 1.0829  & 0.6467  \\
		$C_{3}$   & 1.3376  & 1.3087  & 2.1856 & 1.3116  & 1.3122  & 0.0456 & 1.2868  & 1.2563  & 2.3976  \\
		$C_{4}$   & 1.7613  & 1.6971  & 3.7112 & 1.6932  & 1.7018  & 0.5085 & 1.6381  & 1.5471  & 5.7105  \\
		$C_{5}$   & 2.4973  & 2.3808  & 4.7791 & 2.3445  & 2.3842  & 1.6773 & 2.2399  & 2.0059  & 11.0190 \\
		$\overline{n}$ & 23.6105 & 23.2516 & 1.5318 & 24.4486 & 24.4297 & 0.0775 & 25.5395 & 25.1615 & 1.4913  \\
		\hline
	\end{tabular}

	\begin{tabular}{c|ccc|ccc}
		\hline
		& \multicolumn{3}{c|}{$\sqrt{s}=$ 183 GeV}     & \multicolumn{3}{c}{$\sqrt{s}=$ 189 GeV}      \\
		& Data    & Model   & \% diff. & Data    & Model   & \% diff.  \\
		\hline
		$C_{2}$   & 1.0970  & 1.1008  & 0.3502 & 1.0977  & 1.1004  & 0.2446 \\
		$C_{3}$   & 1.3118  & 1.3324  & 1.5591 & 1.3118  & 1.3265  & 1.1095 \\
		$C_{4}$   & 1.6959  & 1.7705  & 4.3081 & 1.6933  & 1.7462  & 3.0756 \\
		$C_{5}$   & 2.3475  & 2.5737  & 9.1907 & 2.3423  & 2.5018  & 6.5839 \\
		$\overline{n}$ & 26.8559 & 26.8359 & 0.0747 & 26.9417 & 26.7879 & 0.5723 \\
		\hline
	\end{tabular}

\end{table*}

\begin{table*}[]
	\caption{Moments for $p\overline{p}$ fits}
	\label{tab:2b}
	\centering
	\begin{tabular}{c|ccc|ccc|ccc}
		\hline
		& \multicolumn{3}{c|}{$\sqrt{s}=$ 200 GeV}     & \multicolumn{3}{c|}{$\sqrt{s}=$ 546 GeV}     & \multicolumn{3}{c}{$\sqrt{s}=$ 900 GeV}      \\
		& Data    & Model   & \% diff. & Data    & Model   & \% diff. & Data    & Model   & \% diff. \\
		\hline
		$C_{2}$   & 1.2554  & 1.2562  & 0.0674   & 1.2745  & 1.2763  & 0.1449   & 1.3005  & 1.2862  & 1.0991   \\
		$C_{3}$   & 1.8813  & 1.8828  & 0.0799   & 1.9522  & 1.9634  & 0.5708   & 2.0807  & 2.0230  & 2.8112   \\
		$C_{4}$   & 3.2437  & 3.2200  & 0.7350   & 3.4410  & 3.4765  & 1.0279   & 3.8946  & 3.6999  & 5.1265   \\
		$C_{5}$   & 6.2519  & 6.0740  & 2.8865   & 6.7616  & 6.8537  & 1.3524   & 8.2073  & 7.5568  & 8.2527   \\
		$\overline{n}$ & 21.3541 & 21.5906 & 1.1013   & 29.1962 & 29.1271 & 0.2369   & 35.6190 & 35.6841 & 0.1826   \\
		\hline
	\end{tabular}
\end{table*}

\begin{table*}[]
	\caption{Moments for $pp$ fits ($\sqrt{s} =$ 0.9 TeV)}
	\label{tab:2c}
	\centering
	\begin{tabular}{c|ccc|ccc|ccc}
		\hline
		& \multicolumn{3}{c|}{$|\eta|<0.5$}     & \multicolumn{3}{c|}{$|\eta|<1.0$}     & \multicolumn{3}{c}{$|\eta|<1.5$}      \\
		& Data    & Model   & \% diff. & Data    & Model   & \% diff. & Data    & Model   & \% diff. \\
		\hline
		$C_{2}$   & 1.5815  & 1.5808  & 0.0474 & 1.5703  & 1.5429  & 1.7599 & 1.5385  & 1.5256  & 0.8377 \\
		$C_{3}$   & 3.3862  & 3.3965  & 0.3050 & 3.2723  & 3.1682  & 3.2346 & 3.0920  & 3.0439  & 1.5690 \\
		$C_{4}$   & 8.8935  & 8.9659  & 0.8117 & 8.2287  & 7.8883  & 4.2249 & 7.4131  & 7.2363  & 2.4142 \\
		$C_{5}$   & 27.0125 & 27.3465 & 1.2289 & 23.6416 & 22.4624 & 5.1156 & 20.1156 & 19.3766 & 3.7425 \\
		$\overline{n}$ & 4.4260  & 4.4210  & 0.1138 & 8.0591  & 8.1412  & 1.0142 & 11.7442 & 11.7857 & 0.3527 \\
		\hline
	\end{tabular}

	\begin{tabular}{c|ccc|ccc}
	\hline
	& \multicolumn{3}{c|}{$|\eta|<2.0$}     & \multicolumn{3}{c}{$|\eta|<2.4$}      \\
	& Data    & Model   & \% diff. & Data    & Model   & \% diff.  \\
	\hline
	$C_{2}$   & 1.5040  & 1.5011  & 0.1958 & 1.4738  & 1.4739  & 0.0043 \\
	$C_{3}$   & 2.9123  & 2.9093  & 0.1006 & 2.7593  & 2.7587  & 0.0192 \\
	$C_{4}$   & 6.6587  & 6.6529  & 0.0869 & 6.0415  & 6.0207  & 0.3456 \\
	$C_{5}$   & 17.0947 & 17.0034 & 0.5355 & 14.7476 & 14.5723 & 1.1959 \\
	$\overline{n}$ & 15.5073 & 15.4859 & 0.1381 & 18.4315 & 18.4310 & 0.0030 \\
	\hline
	\end{tabular}
\end{table*}

\begin{table*}[]
	\caption{Moments for $pp$ fits ($\sqrt{s} =$ 2.36 TeV)}
	\label{tab:2d}
	\centering
	\begin{tabular}{c|ccc|ccc|ccc}
		\hline
		& \multicolumn{3}{c|}{$|\eta|<0.5$}     & \multicolumn{3}{c|}{$|\eta|<1.0$}     & \multicolumn{3}{c}{$|\eta|<1.5$}      \\
		& Data    & Model   & \% diff. & Data    & Model   & \% diff. & Data    & Model   & \% diff. \\
		\hline
		$C_{2}$   & 1.5848  & 1.5734  & 0.7274 & 1.5657  & 1.5429  & 1.4687 & 1.5229  & 1.4877  & 2.3379 \\
		$C_{3}$   & 3.3344  & 3.2780  & 1.7072 & 3.1870  & 3.1040  & 2.6408 & 2.9579  & 2.8375  & 4.1537 \\
		$C_{4}$   & 8.4428  & 8.1971  & 2.9526 & 7.7060  & 7.4336  & 3.5992 & 6.7493  & 6.4189  & 5.0191 \\
		$C_{5}$   & 24.4903 & 23.4034 & 4.5389 & 21.2185 & 20.2147 & 4.8457 & 17.4755 & 16.5657 & 5.3451 \\
			$\overline{n}$ & 5.2887  & 5.2907  & 0.0391 & 9.7229  & 9.7775  & 0.5601 & 13.9685 & 14.0468 & 0.5584 \\
		\hline
	\end{tabular}
	
	\begin{tabular}{c|ccc|ccc}
		\hline
		& \multicolumn{3}{c|}{$|\eta|<2.0$}     & \multicolumn{3}{c}{$|\eta|<2.4$}      \\
		& Data    & Model   & \% diff. & Data    & Model   & \% diff.  \\
		\hline
		$C_{2}$   & 1.4902  & 1.4656  & 1.6702 & 1.4707  & 1.4685  & 0.1519 \\
		$C_{3}$   & 2.8084  & 2.7311  & 2.7933 & 2.7363  & 2.7418  & 0.1983 \\
		$C_{4}$   & 6.1951  & 6.0161  & 2.9321 & 6.0084  & 6.0757  & 1.1138 \\
		$C_{5}$   & 15.4530 & 15.0953 & 2.3415 & 15.0915 & 15.4570 & 2.3926 \\
		$\overline{n}$ & 18.3244 & 18.2059 & 0.6491 & 21.6960 & 21.4683 & 1.0553 \\
		\hline
	\end{tabular}
\end{table*}

\begin{table*}[]
	\caption{Moments for $pp$ fits ($\sqrt{s} =$ 7 TeV)}
	\label{tab:2e}
	\centering
	\begin{tabular}{c|ccc|ccc|ccc}
		\hline
		& \multicolumn{3}{c|}{$|\eta|<0.5$}     & \multicolumn{3}{c|}{$|\eta|<1.0$}     & \multicolumn{3}{c}{$|\eta|<1.5$}      \\
		& Data    & Model   & \% diff. & Data    & Model   & \% diff. & Data    & Model   & \% diff. \\
		\hline
		$C_{2}$   & 1.7322  & 1.7097  & 1.3103 & 1.7054  & 1.6818  & 1.3928 & 1.6689  & 1.6536  & 0.9216 \\
		$C_{3}$   & 4.1591  & 4.0604  & 2.4030 & 3.9448  & 3.8424  & 2.6298 & 3.7259  & 3.6570  & 1.8662 \\
		$C_{4}$   & 12.2449 & 11.8190 & 3.5396 & 11.0204 & 10.5832 & 4.0474 & 9.9568  & 9.6605  & 3.0209 \\
		$C_{5}$   & 41.4444 & 39.3070 & 5.2939 & 35.0436 & 32.9025 & 6.3022 & 30.1089 & 28.6587 & 4.9354 \\
		$\overline{n}$ & 6.9240  & 6.9915  & 0.9700 & 13.1220 & 13.2729 & 1.1435 & 19.4220 & 19.5972 & 0.8985 \\
		\hline
	\end{tabular}
	
	\begin{tabular}{c|ccc|ccc}
		\hline
		& \multicolumn{3}{c|}{$|\eta|<2.0$}     & \multicolumn{3}{c}{$|\eta|<2.4$}      \\
		& Data    & Model   & \% diff. & Data    & Model   & \% diff.  \\
		\hline
		$C_{2}$   & 1.6402  & 1.6331  & 0.4338 & 1.6161  & 1.6074  & 0.5374 \\
		$C_{3}$   & 3.5647  & 3.5333  & 0.8849 & 3.4326  & 3.3998  & 0.9610 \\
		$C_{4}$   & 9.2037  & 9.0740  & 1.4192 & 8.6134  & 8.5062  & 1.2522 \\
		$C_{5}$   & 26.6864 & 26.0505 & 2.4116 & 24.1689 & 23.7581 & 1.7145 \\
		$\overline{n}$ & 25.8712 & 26.0038 & 0.5111 & 30.7905 & 30.9391 & 0.4815 \\
		\hline
	\end{tabular}
\end{table*}


\section{Discussion}
\label{sec:5}

\subsection{$e^{+}e^{-}$ annihilations}
\label{sec:5.1}

It is immediately apparent that with values of $\chi^{2}$/d.o.f. $\gg 1$ in most of our fits below $\sqrt{s} =$ 133 GeV, our model does not exactly describe $e^{+}e^{-}$ annihilation well.  Considering that the initial interaction is leptonic rather than hadronic, a complete description would certainly require elements outside QCD (e.g. QED) to be built in. Furthermore, $e^{+}e^{-}$ annihilations can result in a variety of final states, and the only one that leads to a scenario described by our model is $e^{+}e^{-} \rightarrow \gamma \rightarrow q\overline{q}$ (electron-position annihilation followed by quark-antiquark pair production). Our model does not account for the other possible interactions.

Considering how the GMD fits excellently to the same set of $e^{+}e^{-}$ data \cite{Dewanto2008}, our model is evidently not a direct extension of the GMD, even though the latter is a partial solution to Giovannini's parton branching equation. However, the fitting results do have an interesting connection – in the GMD, the fitting parameters $k$ and $k'$ are interpreted as the average initial number of quarks and gluons respectively in the branching process \cite{Chan1990}. These are $M$ and $N$ respectively in our formulation, and we can see that unlike hadronic collisions, the branching cascade that results from $e^{+}e^{-}$ annihilation typically begins with many ($>10$) partons, in agreement with [29].

\subsection{$p\overline{p}$ collisions}
\label{sec:5.2}

From Table \ref{tab:1b}, the highest $\chi^{2}$/d.o.f. value obtained is 1.0185, which suggests that our model describes the observed data fairly well. The normalised moments from Table \ref{tab:2b} also suggest good agreement between data and model, except for the fit at $\sqrt{s} =$ 900 GeV, where the model fails to capture the tail of the distribution.

\subsection{$pp$ collisions}
\label{sec:5.3}

From Table \ref{tab:1c}, apart from the fit at $\sqrt{s}=$ 7 TeV and $|\eta| < 1.0$, the $\chi^{2}$/d.o.f. values are generally much smaller than unity. This suggests that our model describes $pp$ collisions at the TeV scale fairly well. Furthermore, the normalised moments agree with each other to within 7.08\%, with the majority within 1\%. However, upon visual inspection, our model does not produce the shoulder-like structure characteristic of KNO scaling violation.

Furthermore, the tails of the produced distributions fall off towards zero faster than the data points. One possible explanation could be that the data points at higher-multiplicity bins are aggregated together in the CMS data, which affected the ability of our optimisation algorithm to reproduce the distribution shape in better detail.

\subsection{The binomial loss function}
\label{sec:5.4}

From Table \ref{tab:1c}, a clear trend for fitted values of the loss parameter $p_{\text{loss}}$ can be observed for $\sqrt{s} =$ 900 GeV and 7 TeV. $p_{\text{loss}}$ increased as the pseudorapidity interval decreased, in agreement with the physical notion behind stage 4 of our model. However, the trend did not show for $\sqrt{s} =$ 2.36 TeV, suggesting that the model is not sufficiently robust.

It must also be noted that the set of values for $p_{\text{loss}}$ for $\sqrt{s} =$ 900 GeV and 7 TeV do not agree with each other exactly, even though the trend within each energy makes sense. This implies that the binomial loss function does not just account for the reduced pseudorapidity space, but gets influenced by other factors as well. Still, it works as a rough approximation to describing data from reduced pseudorapidity space. In future work, the binomial loss function could be improved to become more sophisticated and realistic.

\subsection{Increase in gluonic activity at the TeV scale}
\label{sec:5.5}

Chan and Chew remarked that gluon branching would become increasingly dominant as the centre-of-mass energy $\sqrt{s}$ increases into the TeV regime \cite{Chan1990}. In the context of the GMD, this would manifest itself via the parameters $k$ approaching zero and $k'$ increasing above 3.0, which turns the GMD into the Furry-Yule distribution (FYD, which is another special case of the GMD where $k = 0$). We observe these predictions validated in our results – from Table \ref{tab:1c}, the fitted values of $\widetilde{A}$ decreases as $\sqrt{s}$ increases from 0.9 to 7 TeV, which is characteristic of a Furry-Yule process. Also, the fraction of initial partons being gluons also increases, and the values of $C$ (probability of $g \rightarrow ggg$) increases. Together, our data strongly agrees with their prediction.

It must be noted that the best-fit parameters presented in section \ref{sec:4} are by no means representative of the absolute global minimum values for $\chi^{2}$. The parameter space is infinitely dense to be sampled thoroughly; instead, they are the best-possible result found using an optimisation heuristic within available computational resources. Since $\chi^{2}$ in our model is a function of 8 to 9 parameters, it is entirely conceivable that distinct parameter sets could result in very comparable $\chi^{2}$ values. In particular, the multiplicity distributions published by experiments are a result of averaging over numerous events, and are certainly not describable by a fixed integer value of $M$ and $N$.

\section{Prediction for $\sqrt{s} =$ 14 TeV}
\label{sec:6}

Based on the best-fit values summarised in Table \ref{tab:1c}, we attempt to make a prediction for $pp$ collisions at $\sqrt{s} =$ 14 TeV. We continue to expect gluonic activity to be dominant, which would manifest itself in the following ways:

\begin{enumerate}
	\item Processes $A$, $B$ and $C$ would become even more prominent, and hence their values would increase.
	\item Process $\widetilde{A}$ would decrease further towards zero.
	\item 	Each parton would hadronise into more hadrons, and hence $l_{\text{avg}}$ and $l_{\text{max}}$ would increase above 3.
	\item The initial conditions would still be best described by the presence of only gluons and no quarks.
	\item The overall probability of gluon branching $A+B+C$ appears to increase with $\sqrt{s}$. However, at $\sqrt{s} =$ 7 TeV, $A+B+C = 0.8913$ which is nearing the upper bound of 1 (conservation of probability). Hence, the value of $T$ would need to increase beyond 3.0 for the model to describe the final multiplicity distribution for $\sqrt{s} =$ 14 TeV (see comments in section \ref{sec:2.1}, stage 1).
	
\end{enumerate}

\section{Conclusion}
\label{sec:7}

A numerical model for producing charged-particle multiplicity distributions has been presented. It has been shown to describe hadronic collision data well up to the TeV scale, but less so for lower-energy leptonic interactions. This work has also shown that within the framework of pQCD and multiparticle production, Giovannini's parton branching equations remain valid up to the TeV scales currently being explored at the LHC. Our model suggests that the increase in gluonic activity as $\sqrt{s}$ increases in hadronic collisions can be attributed to a higher proportion to gluon branching and initial gluon number, in agreement with earlier predictions by Chan and Chew \cite{Chan1990}.

\begin{acknowledgements}
The authors would like to thank the National University of Singapore, and the support and helpful discussions with colleagues. This work is supported by the NUS Research scholarship.
\end{acknowledgements}

\bibliographystyle{spphys}
\bibliography{References.bib}   

\end{document}